\documentclass[12pt]{iopart}
\usepackage{iopams} 
\usepackage{graphicx} 
\begin{document}

\title{`17-15' Superconductivity - Rh$_{17}$S$_{15}$ and Pd$_{17}$Se$_{15}$}

\author{Naren HR$^1$, Arumugam Thamizhavel$^1$, Sushil Auluck$^2$, Rajendra Prasad$^2$ and Ramkrishnan S$^1$}

\address{$^1$Tata Institute of Fundamental Research, Mumbai, India}
\address{$^2$Indian Institute of Technology, Kanpur, India}
\ead{nareni@tifr.res.in}

\begin{abstract}
The presence of strongly correlated superconductivity in Rh$_{17}$S$_{15}$ has been recently established. In this work, we compare the normal and superconducting parameters of a single crystal of Rh$_{17}$S$_{15}$ with those of a polycrystalline sample of Pd$_{17}$Se$_{15}$ which is reported here for the first time to be a superconductor below 2.2 K. Pd$_{17}$Se$_{15}$, which is iso-structural to Rh$_{17}$S$_{15}$ (space group Pm3m), has very different properties and provides for an interesting study in contrast with Rh$_{17}$S$_{15}$. We see that large unit volume of Pd$_{17}$Se$_{15}$ and the large separation of Pd-Pd atoms in its structure as compared those in a pure Pd metal lead to the absence of strong correlations in this compound. Finally, we report band structure calculations on these two compounds and compare the density of states with estimates from heat capacity data. Upper critical field, Heat capacity (down to 500 mK), Hall effect and band structure studies suggest that Rh$_{17}$S$_{15}$ is a multiband superconductor. 
\end{abstract}


\maketitle

\section{Introduction}
Recently there has been intense activity in the study of transition metal sulphides and selenides due to a variety of their properties such as, strongly correlated superconductivity, spin and charge density waves. In our effort to understand the physical properties of noble metal chalcogenides, we have studied cubic Rh$_{17}$S$_{15}$ (Ref.~\cite{r1,r2}) which is considered as an excellent catalyst by chemists (Ref.~\cite{r3,r4}) and Pd$_{17}$Se$_{15}$.

\begin{figure*}
	\centering
		\includegraphics[width=12cm]{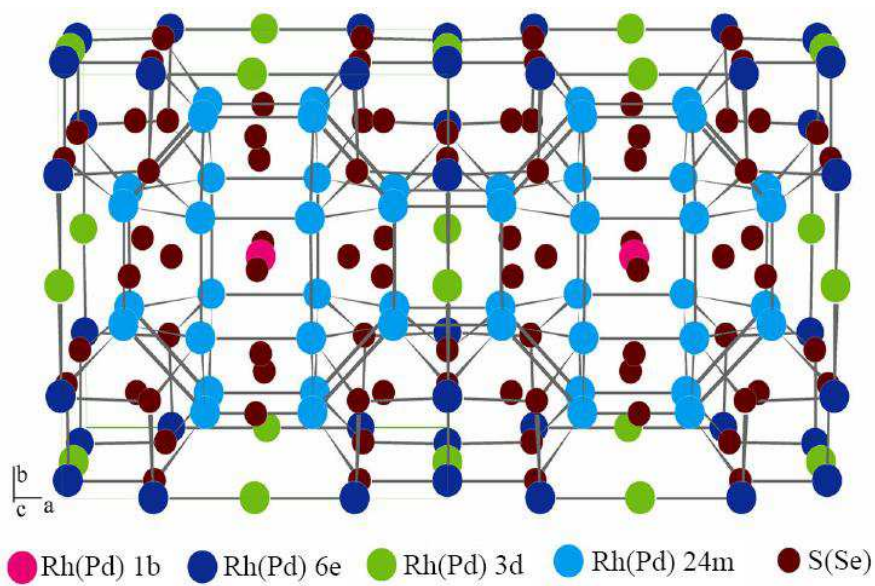}
	\caption{{[colour online] The crystal structure of Rh$_{17}$S$_{15}$ and Pd$_{17}$Se$_{15}$ contains two formula units per unit cell}}
	\label{struct}
\end{figure*}

Rh$_{17}$S$_{15}$ (known as Miassite), a mineral mainly found near the Miass river in Russia, and  Pd$_{17}$Se$_{15}$ (known as Palladseite), a mineral mainly found in Brazil, are iso-structural cubic compounds whose crystal structure is shown in fig. \ref{struct}. They have a large unit cell with 64 atoms in each (two formula units) and Rh (or Pd) occur in four inequivalent crystallographic sites. Some of us had reported strongly correlated superconductivity (T$_{c}$ = 5.4 K) in Rh$_{17}$S$_{15}$ (Ref.~\cite{r1}) with magnetization, resistivity, heat capacity and Hall measurements. The compound was remarkable for its enhanced susceptibility, large Sommerfeld coefficient (108 mJ/mol-K) and unusually high upper critical field (H$_{c2}$) values (around 13 T at 3 K) well above the Pauli limiting field (9.9 T). We had ascribed the origin of strong correlations to the possible occurrence of a narrow Rh d-band of states at the Fermi level possibly due to the close lying Rh atoms at the 3d and 6e positions. The distance between these atoms (2.58 \r{A}) is even lesser than that between the closest Rh atoms in Rh metal (2.69 \r{A}). In iso-structural Pd$_{17}$Se$_{15}$, however, the lowest Pd-Pd distance (2.78 \r{A}) is more than that in elemental Pd (2.75 \r{A}). Pd$_{17}$Se$_{15}$ is found to be a superconductor below 2.2 K. 

Our paper is arranged as follows. The first section deals with the sample growth and structural characterisation details. The second has results and discussion on the superconducting properties of Rh$_{17}$S$_{15}$. The third deals with the superconducting properties of Pd$_{17}$Se$_{15}$ which are compared with those of Rh$_{17}$S$_{15}$. In the fourth section we report band structure calculations on these systems. The fifth section contains the conclusions.

\section{The Samples}
The compounds were prepared by heating stoichiometric compositions of the two elements at the rate of 8 deg/hr to a temperature of 1150 deg C in an alumina crucible that was sealed in an evacuated quartz tube. Thus reacted mixture is annealed at 1050 deg C for two days and then cooled down firstly at 8 deg/hr till 600 deg C and further at 30 deg/hr. The samples were characterised using powder XRD and the single crystal of  Rh$_{17}$S$_{15}$ showed clear cubic spots of Pm3m space group in a Laue diffractometer. The lattice constants were estimated to be around 9.9093(2) \r{A} for Rh$_{17}$S$_{15}$ after a Reitveld analysis. 

\begin{figure*}
	\centering
		\includegraphics[width=12cm]{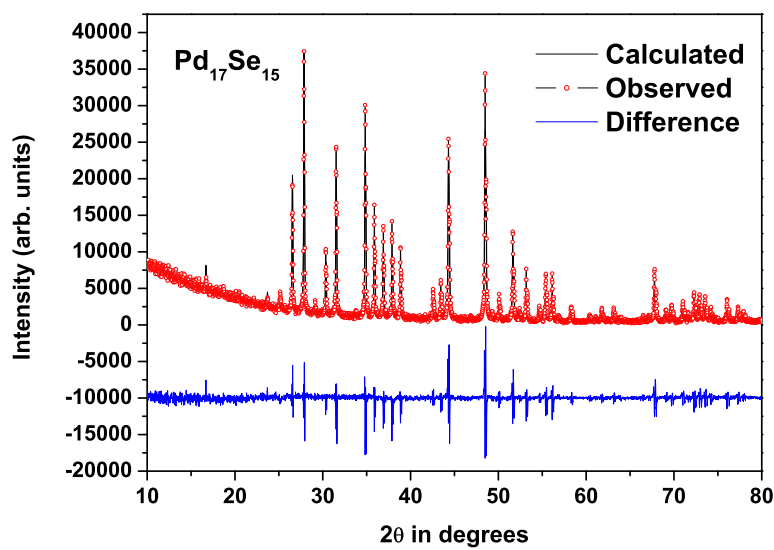}
	\caption{{[colour online] The powder XRD pattern of Pd$_{17}$Se$_{15}$ as refined with the calculated Pm3m structure of Pd$_{17}$Se$_{15}$ by a Reitveld analysis}}
	\label{xrd}
\end{figure*}

Since this is the first report on superconductivity in Pd$_{17}$Se$_{15}$, we have confirmed that the compound is indeed in a single phase by both powder XRD and Electron Probe Micro Analysis (EPMA). In fig. \ref{xrd} we show the refined XRD pattern for Pd$_{17}$Se$_{15}$. There is no impurity peak and the pattern refines to the cubic Pm3m structure of Pd$_{17}$Se$_{15}$ with a lattice constant of 10.6044(7) \r{A} by a Reitveld analysis. Hence, the XRD pattern indicates an essentially single phase sample. Phase sensitive EPMA shows almost uniform stoichiometry with Pd:Se concentration of 52.25:47.75 atomic percent in the analysed region (an area of 1.5mm x 2mm). A 17:15 ratio of Pd and Se should give a 53.1:46.9 atomic percent ratio. Providing for a small error in estimation, we conclude that the majority of the sample has indeed formed in the desired 17:15 ratio. However, small regions of Pd:Se concentration of 49.62:50.38 atomic percent are seen. Inspite of small off-stoichiometric regions, our results indicate that there is indeed bulk superconductivity in Pd$_{17}$Se$_{15}$. 

The specific heat capacity measurements were performed in a Physical property measurement system (Quantum Design, USA). The magnetisation measurements were done in a Magnetic property measurement system (Quantum Design, USA). Transport measurements were performed in a home-made system and in an Adiabatic demagnetisation refrigerator (Cambridge magnetic refrigeraton, UK).

\section{Superconducting Properties of single crystalline R{\lowercase{h}}$_{17}$S$_{15}$}

\subsection{Specific Heat Capacity measurements}
\begin{figure*}
	\centering
		\includegraphics[width=12cm]{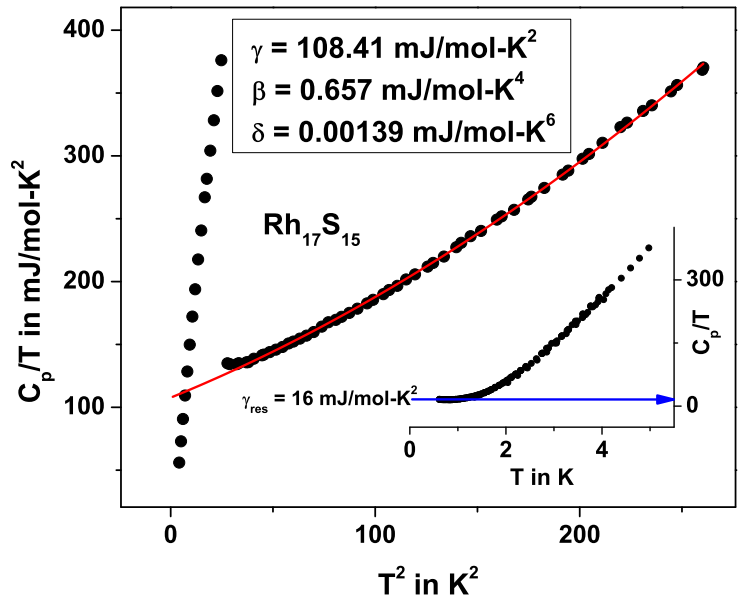}
	\caption{{[colour online] C$_{p}$/T versus T$^{2}$ for Rh$_{17}$S$_{15}$. The continuous line is a fit to the equation mentioned in the text. Inset shows the low temperature C$_{p}$/T versus T data which levels off at $\gamma_{res}\approx$ 16 mJ/mol-K$^{2}$}}
	\label{RhSCp}
\end{figure*}

In fig. \ref{RhSCp} we show the zero field specific heat capacity (C$_{p}$) data with C$_{p}$/T plotted against T$^{2}$. The continuous line is a fit to 

$$ C_p~=~ \gamma~T+~\beta~T^{3}~+~\delta~T^5 \eqno(1) $$

in the low temperature region in the normal state where $\gamma$ is the electronic contribution, $\beta$ is the lattice contribution and $\delta$ is the contribution due to anharmonicity. We obtain a $\gamma$ of 108.41 mJ/mol-K$^{2}$, $\beta$ of  0.657 mJ/mol-K$^{4}$ and $\delta$ of  0.00139 mJ/mol-K$^{6}$. The presence of a significant anharmonicity could be a result of a complex phonon structure in these compounds which could arise largely due to a complex crystal structure. Using an equal entropy estimation, we can estimate a T$_{c}$ of 5.36 K. From $\beta$ we obtain a Debye temperature ($\theta$$_{D}$) of 456 K which indicates a hard lattice. Using $\theta$$_{D}$ and McMillan's equation (Ref.~\cite{r5}) with a Coulomb parameter ($\mu$* = 0.13) we obtain an electron-phonon coupling constant ($\lambda$) of 0.58 which puts Rh$_{17}$S$_{15}$ in the intermediate coupling regime. However, from the jump at the transition (around 1558.554 mJ/mol-K), $\Delta C$$_{p}$/$\gamma$ T$_{c}$ is around 2.68 which indicates that the electron-electron interaction is in the strong-coupling regime as opposed to a weak-coupling BCS value of 1.43. Therefore, we conclude that phonon-mediated mechanism is insufficient to account for strong electron correlations in Rh$_{17}$S$_{15}$.
Using the value of $\gamma$ one can also compute the enhanced density of states N*(0) from N*(0) = 0.4248 $\gamma$ states/eV-formula unit where $\gamma$ is in mJ/mol-K$^{2}$. With the earlier stated $\gamma$ value we get N*(0) = 46 states/eV-formula unit.
 
The inset in fig. \ref{RhSCp} shows very low temperature heat capacity data with C$_{p}$/T plotted against T. Clearly the low temperature values of C$_{p}$/T seem to level off at a residual non-zero value ($\gamma_{res}\approx$ 16 mJ/mol-K$^{2}$) whereas we expect the heat capacity to go to zero as the temperature approaches 0K. This behaviour has been seen in multi-band superconductors like MgB$_{2}$ and more recently in the pnictides (Ref.~\cite{r19}). Such systems typically have multiple gaps in the superconducting state where some gap is small enough to contribute to the heat capacity even at low temperatures. 

\subsection{Upper Critical Field measurements}
\begin{figure*}
	\centering
		\includegraphics[width=12cm]{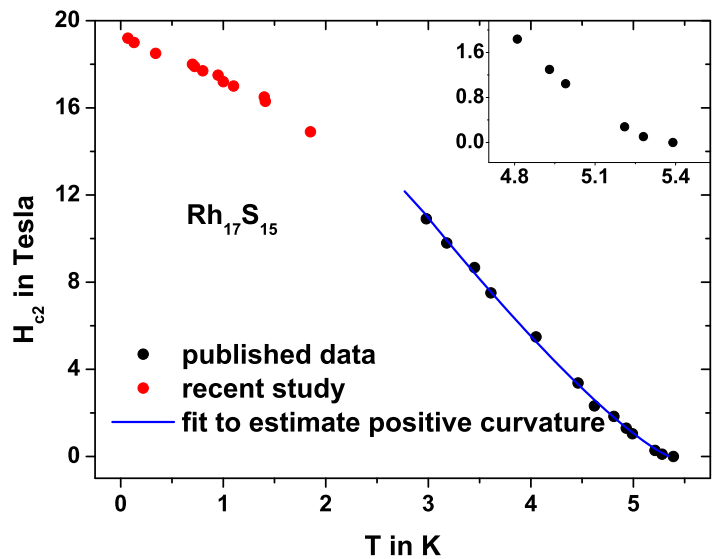}
	\caption{{[colour online] H$_{c2}$ versus temperature for Rh$_{17}$S$_{15}$. The inset shows the positive curvature at the transition. The continuous line is a fit to the equation mentioned in the text}}
	\label{RhSHc2}
\end{figure*}

Fig. \ref{RhSHc2} shows the upper critical field of Rh$_{17}$S$_{15}$ as a function of temperature from 70 mK to 5.4 K as estimated from ac and dc magnetization measurements. The very low temperature data (below 2 K) have been included from a recent study (Ref.~\cite{r6}). From this one can estimate the value of the upper critical field at H$_{c2}$ at 0 K to be 20 T which is in
agreement with the recent single crystal study of Settai et al (Ref.~\cite{r7}). The slope (dH$_{c2}$/dT) near the superconducting transition is estimated to be - 3.6 T/K which is smaller than the value quoted by Settai et al (Ref.~\cite{r7}). From H$_{c2}$(0), we can estimate the BCS coherence length ($\xi$$_{0}$) using H$_{c2}(0)$ = $\phi$$_{0}$/2$\pi$$\xi$$_{0} ^{2}$ which gives us a  $\xi$$_{0}$ of 4*10$^{-7}$ cm. 

A strong-coupling BCS theory estimation of the orbital limit of the upper critical field of a type II superconductor is the WHH theory expression (Ref.~\cite{r8})
 
$$H_{c2}(0) = - 0.69T_{c}\left[\frac{dH_{c2}}{dT}\right]_{T_{c}} \eqno{2}$$

from which we get H$_{c2}$(0) to be 13.3 T which is much smaller than the actual value. 

The Pauli paramagnetic limiting field (H$_{P}$(0)), which refers to the energy scale required to depair the Cooper-pairs, is roughly 1.84*T$_{c}$ (in Tesla) for a BCS superconductor. This value is around 9.9 T for Rh$_{17}$S$_{15}$ which is half the actual value of H$_{c2}$(0). Taking into account the electron phonon coupling ($\lambda$) we can estimate an enhanced Pauli limit (Ref.~\cite{r9}) H$_{P}$(0)* = 1.84T$_{c}$(1+$\lambda$)$^{1/2}$ which gives a value of  12.4 T for Rh$_{17}$S$_{15}$ which is again much smaller than the actual value. For Rh$_{17}$S$_{15}$ to be Pauli limited in a strong coupling BCS sense, the coupling parameter $\lambda$  will have to be much larger than 3. 
It can be speculated that the reason for such a large upper critical field could be strong electron-electron correlations (as in the heavy fermions) or unconventional pairing or large spin-orbit coupling (as in the Chevrel Phases) or a combination of the three. We cannot conclusively decide on any one reason based on the available data.
We have also estimated the penetration depth $\lambda$$_{0}$ as 7000 \r{A} from the $\mu$SR measurements performed at PSI, Switzerland recently (Ref.~\cite{r10}). Using $\lambda$$_{0}$ and $\xi$$_{0}$ we estimate the Ginzburg Landau parameter $\kappa$ to be around 175 which makes Rh$_{17}$S$_{15}$ an extreme type II superconductor. From this one can estimate a lower critical field of 16 Oe using H$_{c1}(0)$ = $\phi$$_{0}$ln$\kappa$/4$\pi$$\xi$$_{0} ^{2}$.

We also note that the positive curvature of the H$_{c2}$ curve near T$_{c}$ is typical of multi-band superconductors like MgB$_{2}$ (Ref.~\cite{r11}) and the Borocarbides (Ref.~\cite{r12}). Using the emperical expression 

$$H_{c2} = H_{c2}^*(1-T/T_{c})^{1+a}\eqno{3}$$

for quantifying this positive curvature (Ref.~\cite{r11, r12}), we get a good fit from 0.5T$_{c}$\textless T\textless T$_{c}$ as seen in fig. \ref{RhSHc2} with a H$_{c2}^{*}$ of 30.3 T and `a' of 0.24 (a value similar to Borocarbides and MgB$_{2}$  in clean limit). This, the low temperature heat capacity data (previous subsection) and the fact that the Hall data in Rh$_{17}$S$_{15}$ shows a sign change in majority carriers (Ref.~\cite{r1}) strongly hint at a multi band scenario for Rh$_{17}$S$_{15}$.

\section{Superconducting properties of polycrystalline P\lowercase{d}$_{17}$S\lowercase{e}$_{15}$}

\subsection{Resistivity measurements}
\begin{figure*}
	\centering
		\includegraphics[width=12cm]{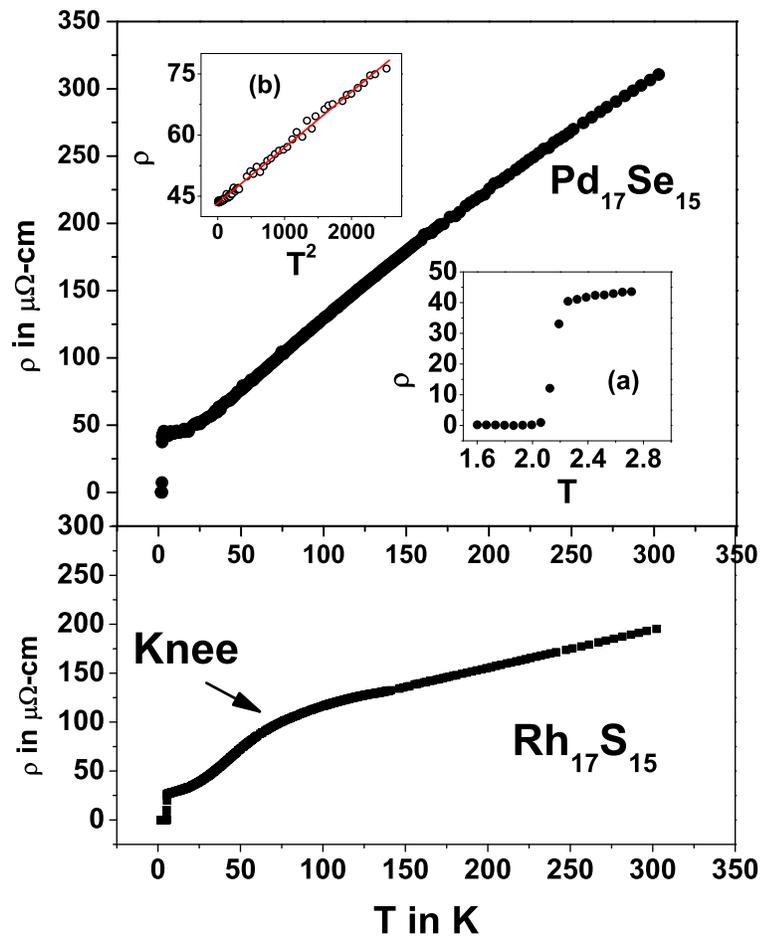}
	\caption{{[colour online] Resistivity data of Pd$_{17}$Se$_{15}$ and Rh$_{17}$S$_{15}$. Inset (a) shows the transition region and inset (b) shows a T$^{2}$ fit. The lower figure is the Rh$_{17}$S$_{15}$ curve showing the knee-like feature}}
	\label{Rhos}
\end{figure*}

Fig. \ref{Rhos} shows the temperature dependence of the resistivity ($\rho$(T)) of Pd$_{17}$Se$_{15}$ from 1.6 to 300 K. Inset (a) shows the transition region (T$_{c}$ = 2.2 K for Pd$_{17}$Se$_{15}$). 
Inset (b) shows a T$^{2}$ fit to the low temperature data ($\rho$ = $\rho$$_{0}$ + A*T$^{2}$) which gives a value of 43.2 $\mu$$\Omega$cm for the residual resistivity ($\rho$$_{0}$) and a value of 0.014 $\mu\Omega$ cm/K$^{2}$ for A. The value of A for Pd$_{17}$Se$_{15}$ is comparable to that of Rh$_{17}$S$_{15}$ Ref.~\cite{r1}. 

The lower graph in fig. \ref{Rhos} shows the resistivity data of the single crystal of Rh$_{17}$S$_{15}$ for comparision. The two curves are distinctly different in the appearance of a knee-like feature at around 60 K in Rh$_{17}$S$_{15}$ which is completely absent in Pd$_{17}$Se$_{15}$ which has a conventional metallic behaviour. We do not, at present, understand the reason for the knee-like feature. However we note that this kind of behaviour has been observed in many transition metal based superconductors, particularly in many A-15 compounds (Ref.~\cite{r13}) which have a sharp feature in density of states at Fermi level. In that scenario, the Fermi level moves to drastically lower values of density of states at higher temperatures and causes a knee in the resistivity. Such a mechanism could be in operation in Rh$_{17}$S$_{15}$. We also note that the sign change in majority carriers as reported in (Ref.~\cite{r1}) occurs at around 60 K as well.

\subsection{Specific Heat Capacity measurements}
\begin{figure*}
	\centering
		\includegraphics[width=12cm]{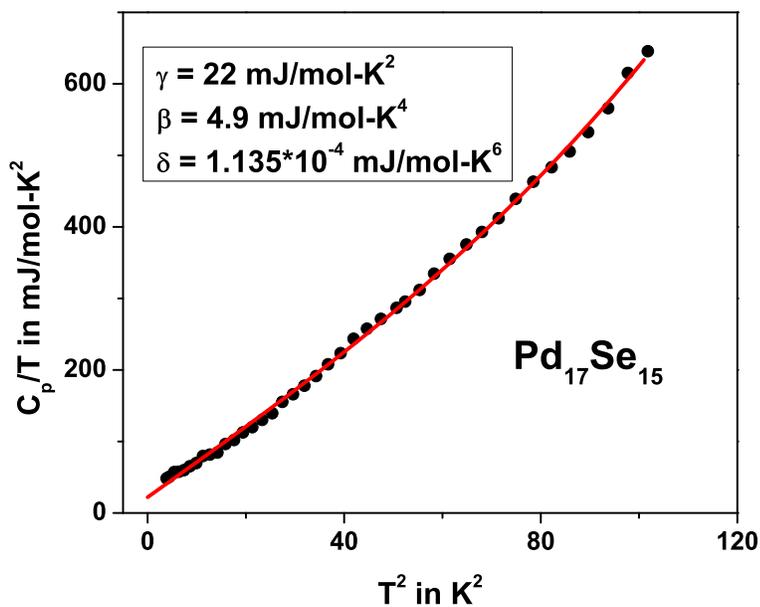}
	\caption{{[colour online] C$_{p}$/T versus T$^{2}$ for Pd$_{17}$Se$_{15}$. The continuous line is a fit to the equation mentioned in the text}}
	\label{PdSeCp}
\end{figure*}

\begin{figure*}
	\centering
		\includegraphics[width=12cm]{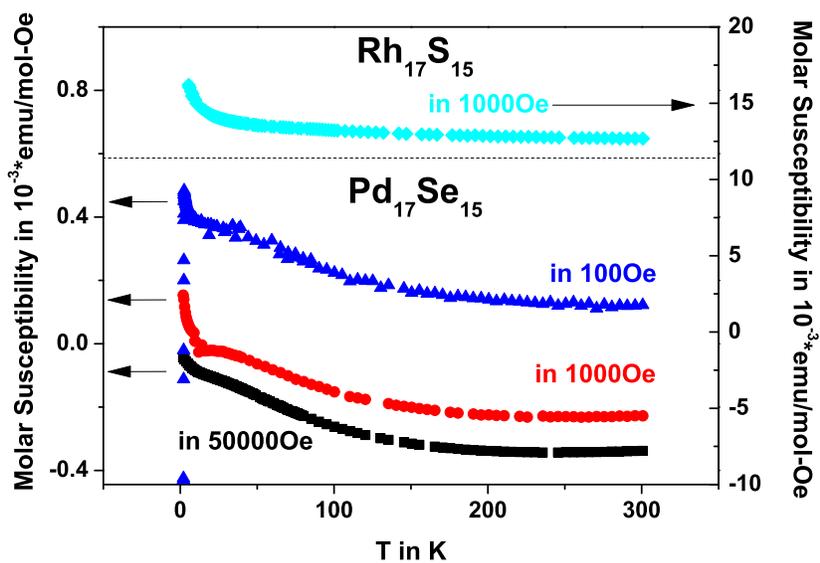}
	\caption{{[colour online] DC susceptibility curves of Rh$_{17}$S$_{15}$ and Pd$_{17}$Se$_{15}$ at different fields. Note the two different scales for the two compounds}}
	\label{dcmags}
\end{figure*}

Fig. \ref{PdSeCp} shows a plot of C$_{p}$ versus T for the polycrystalline Pd$_{17}$Se$_{15}$ in the absence of applied magnetic field from 2 to 15 K. Our range of C$_{p}$ measurement, unfortunately cannot capture the transition region well. The solid line in the same figure is a fit to equation (1) described earlier. The fit yields a $\gamma$ of 22 mJ/mol-K$^{2}$, $\beta$ of 4.9 mJ/mol-K$^{4}$ and  of 0.11 $\mu$J/mol-K$^{6}$. From the value of $\beta$ we obtain a Debye temperature ($\theta$$_{D}$) of 233.3 K which indicates that Pd$_{17}$Se$_{15}$ is a much softer lattice as compared with Rh$_{17}$S$_{15}$. From the values of T$_{c}$ and T$_{c}$, we estimate a value of 0.54 for the electron-phonon coupling constant (Ref.~\cite{r5}) for Pd$_{17}$Se$_{15}$ which is very similar to the  value of Rh$_{17}$S$_{15}$. The anharmonicity is smaller than that of Rh$_{17}$S$_{15}$. More importantly, the enhanced density of states, N*(0), as calculated from the earlier expression, is around 9.35 states/eV-formula unit and is smaller than that of Rh$_{17}$S$_{15}$ by a factor of 5 indicating that the correlations are much weaker in Pd$_{17}$Se$_{15}$. The reduction in the density of states in Pd$_{17}$Se$_{15}$ as compared to that of Rh$_{17}$S$_{15}$ is also reflected in the temperature dependence of normal state magnetic susceptibility as shown in the fig. \ref{dcmags}.

\subsection{DC magnetisation measurements}
Fig. \ref{dcmags} shows the dc susceptibility curves of Rh$_{17}$S$_{15}$ and Pd$_{17}$Se$_{15}$ at different fields. Since none of Rh, Pd, S or Se carry a magnetic moment we expect to see a temperature independent susceptibility, essentially sum of Pauli paramagnetism, Landau diamagnetism and core diamagnetism. However, although the values of susceptibility ($\chi$(T)) are small, the data display a distinct temperature dependence for Rh$_{17}$S$_{15}$. There is a gradual increase in susceptibility as one goes to lower temperatures until the superconducting transition. The weak temperature dependence of $\chi$(T) could arise due to a sharp density of states at the Fermi level which leads to the temperature dependence of the Pauli spin susceptibility. We believe that the mechanism is similar to the observed in the A-15 compound, V$_{3}$Si (Ref.~\cite{r14}). This conjecture is recently supported by the observation of temperature dependent Knight shift from the 103Rh-NMR experiment Ref.~\cite{r15}. However the $\chi$(T) of Pd$_{17}$Se$_{15}$ is diamagnetic with impurities (presumably at few ppm levels) dominating at low fields. The value of $\chi$(T) is nearly two orders smaller than that of Rh$_{17}$S$_{15}$ indicating that the density states is much smaller which is confirmed by the heat capacity data.

\subsection{Upper Critical Field measurements}
\begin{figure*}
	\centering
		\includegraphics[width=12cm]{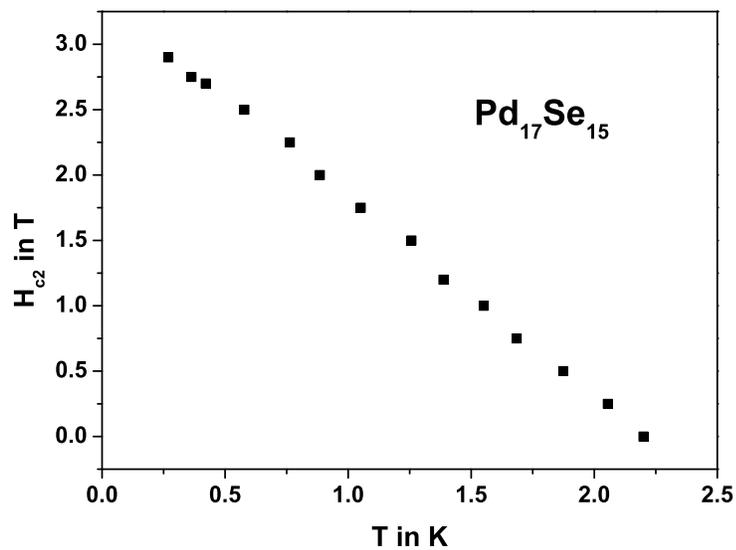}
	\caption{{[colour online] H$_{c2}$ versus temperature for Pd$_{17}$Se$_{15}$}}
	\label{PdSeHc2}
\end{figure*}

Finally, we show the temperature dependence of the upper critical field (H$_{c2}$(T)) of Pd$_{17}$Se$_{15}$ from 270 mK to 2.2 K in Fig. \ref{PdSeHc2}. The curvature of H$_{c2}$ nearT$_{c}$ is more like a conventional superconductor as opposed the one seen in Rh$_{17}$S$_{15}$. However, both show a tendency to saturate at lower temperatures. The saturation is better seen in Rh$_{17}$S$_{15}$ because we have data upto 70 mK whereas, we have data only upto 270 mK in Pd$_{17}$Se$_{15}$. The H$_{c2}$(0) is estimated to be 3.3 T by an extrapolation which is nearly 6 times smaller than that of Rh$_{17}$S$_{15}$. The BCS coherence length $\xi$$_{0}$ is 100 \r{A} as estimated from the H$_{c2}$(0). The slope (dH$_{c2}$/dT) near T$_{c}$ is -1.5 T/K which yields a value of 2.3 T for H$_{c2}$(0) from WHH theory (equation (2)). However, the Pauli paramagnetic limiting field (H$_{P}$(0)) is estimated to have a value of 4 T which suggests that that orbital critical field is Pauli limited in contrast to the situation in Rh$_{17}$S$_{15}$. The upper critical field data indicate the Pd$_{17}$Se$_{15}$ is a conventional superconductor as opposed to Rh$_{17}$S$_{15}$.

\section{Band Structure Calculations}
In this section, we present some band structure calculations on these two compounds. Firstly, we will present a LDA (Local Density Approximation) calculation of the density of states and band structure and compare with the experimental data. Later we will present an LDA+U (which takes into account Coulombic interaction between electrons) calculation and compare that with experiments.

\subsection{LDA calculations}
These calculations were performed using the WIEN2K code (Ref.~\cite{r16}) which is a Full Potential Linearised Augmented Plane Wave calculation (FPLAPW). Calculations were also performed with a Full Potential Local Orbital code (FPLO) (Ref.~\cite{r17,r18}) and the results are very similar to the WIEN2K results. Also the above calculations were performed in an LDA framework as well as in a Generalised Gradient Approximation framework and the results match very closely. We will be presenting the WIEN2K calculations in an LDA framework here.

\begin{figure*}
	\centering
		\includegraphics[width=12cm]{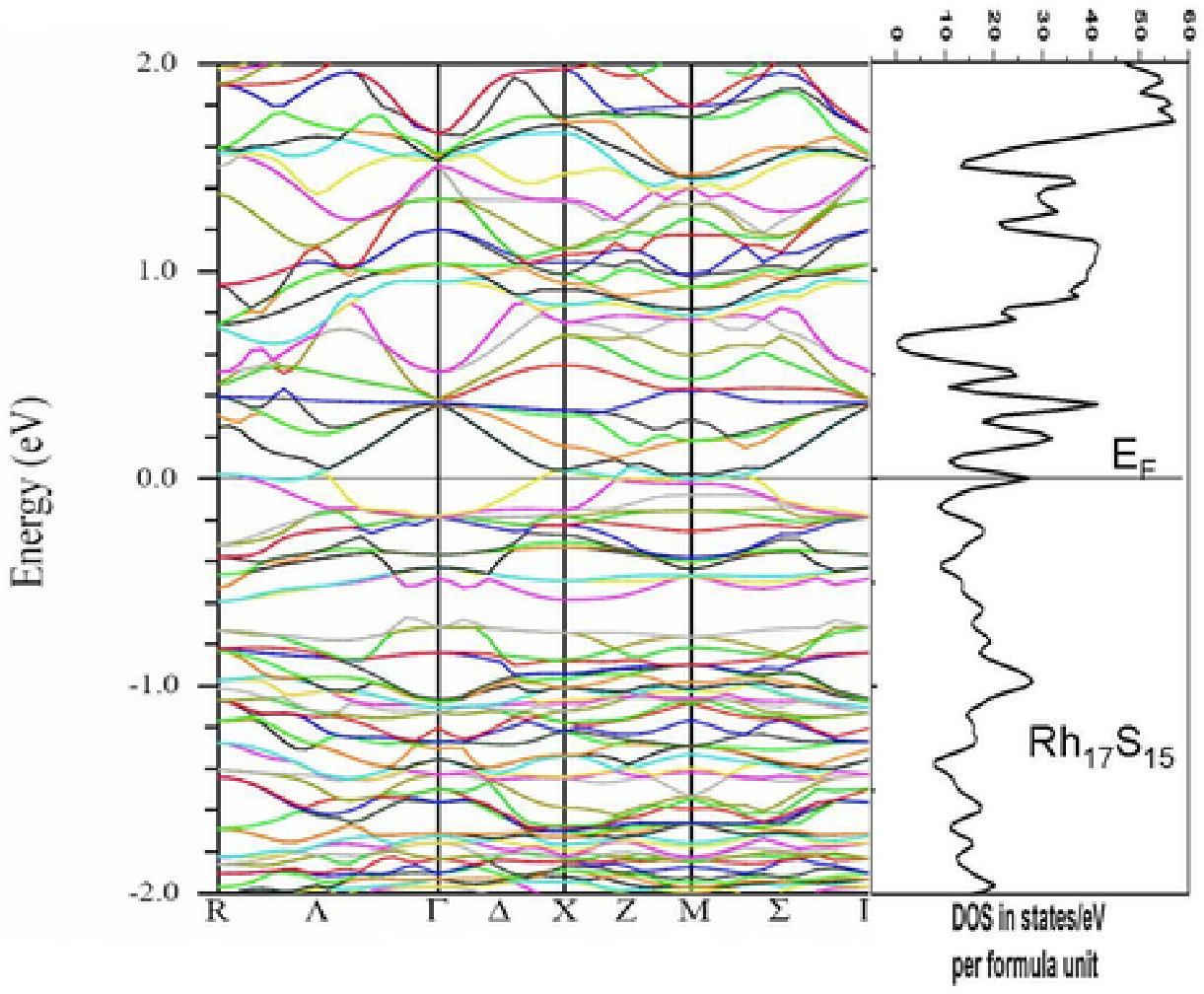}
	\caption{{[colour online] The band structure of Rh$_{17}$S$_{15}$. The total DOS is indicated on the right side of the plot}}
	\label{RhSbs}
\end{figure*}

In fig. \ref{RhSbs} we have the band structure of Rh$_{17}$S$_{15}$ juxtaposed with the Density of States plot. We see that the Fermi level (here E$_{F}$ = 0) indeed lies on a peak of the DOS. We can see both electron-like and hole-like pockets at the Fermi Level supporting the fact that there are both types of charge carriers in the system. This also supports the conjecture that Rh$_{17}$S$_{15}$ could have a strong compensation due to oppositely charged carriers as speculated by Settai et al from the magnetoresistance data (Ref.~\cite{r7}). Around the M direction, we also see flat electron-like and hole-like bands which could contribute to the peak in the DOS at Fermi level. Such flat bands also hint at a multi-band scenario supporting the Hall data and the fact that the upper critical field curve has a multi-band like positive curvature.  

\begin{figure*}
	\centering
		\includegraphics[width=12cm]{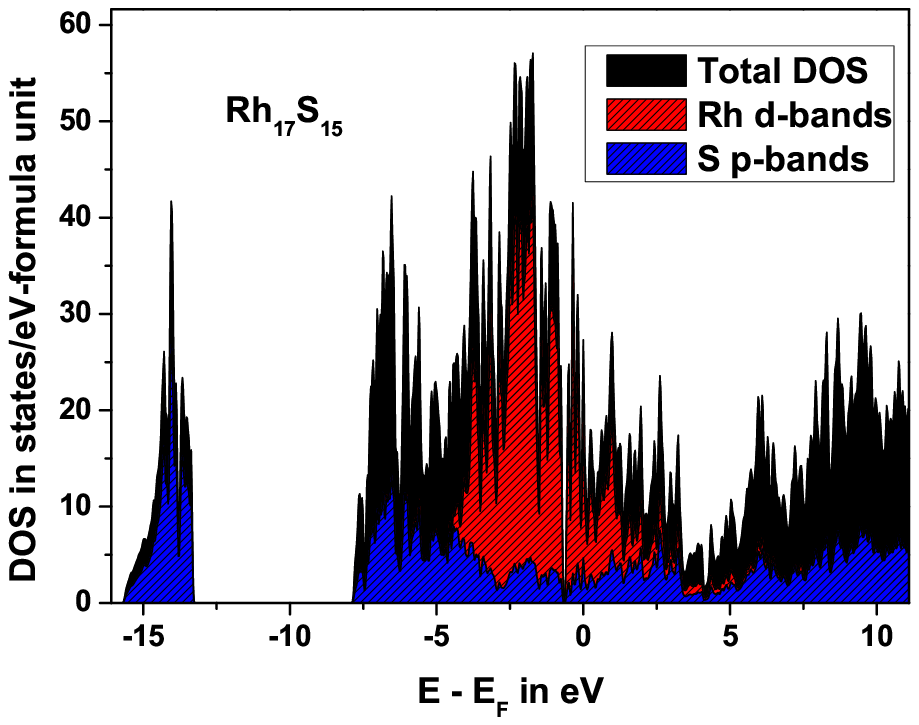}
	\caption{{[colour online] DOS for Rh$_{17}$S$_{15}$ from LDA calculations. Contributions of Rh and S are indicated in red and blue respectively}}
	\label{RhSDOS}
\end{figure*}

Fig. \ref{RhSDOS} shows the DOS plot in the total energy range of calculation. The DOS at Fermi level is around 27 states/eV-formula unit. We recall that the ehanced DOS from the heat capacity data was around 46 states/eV-formula unit. From this one can estimate the electron-phonon coupling ($\lambda$) using 

$$\lambda = (\frac{\mathrm{DOS~from~heat~capacity}}{\mathrm{DOS~from~band~structure}})- 1 \eqno{4}$$

which gives a $\lambda$ of 0.7. This value is slightly larger than that calculated from Mcmillan's formula (0.58). 

In fig. \ref{RhSDOS}, we have also indicated the contributions by Rhodium d-bands and Sulphur p-bands to the DOS. Nearly 80 percent of the total DOS is contributed by the Rh d bands of which the largest contributors are the Rh24m atoms mainly because of the large number of such atoms. Interestingly, when one calculates a contribution per atom, it turns out that the Rh24m atoms still contribute more than twice the other Rh or S atoms. This is in contrast with our earlier speculation that it must be the closer lying Rh3d and Rh6e atoms that contribute to the DOS at Fermi level. However, this supports the NMR data which says that that the Rh 24m atoms contribute the maximum towards the Knight shift (Ref.~\cite{r15}).

\begin{figure*}
	\centering
		\includegraphics[width=12cm]{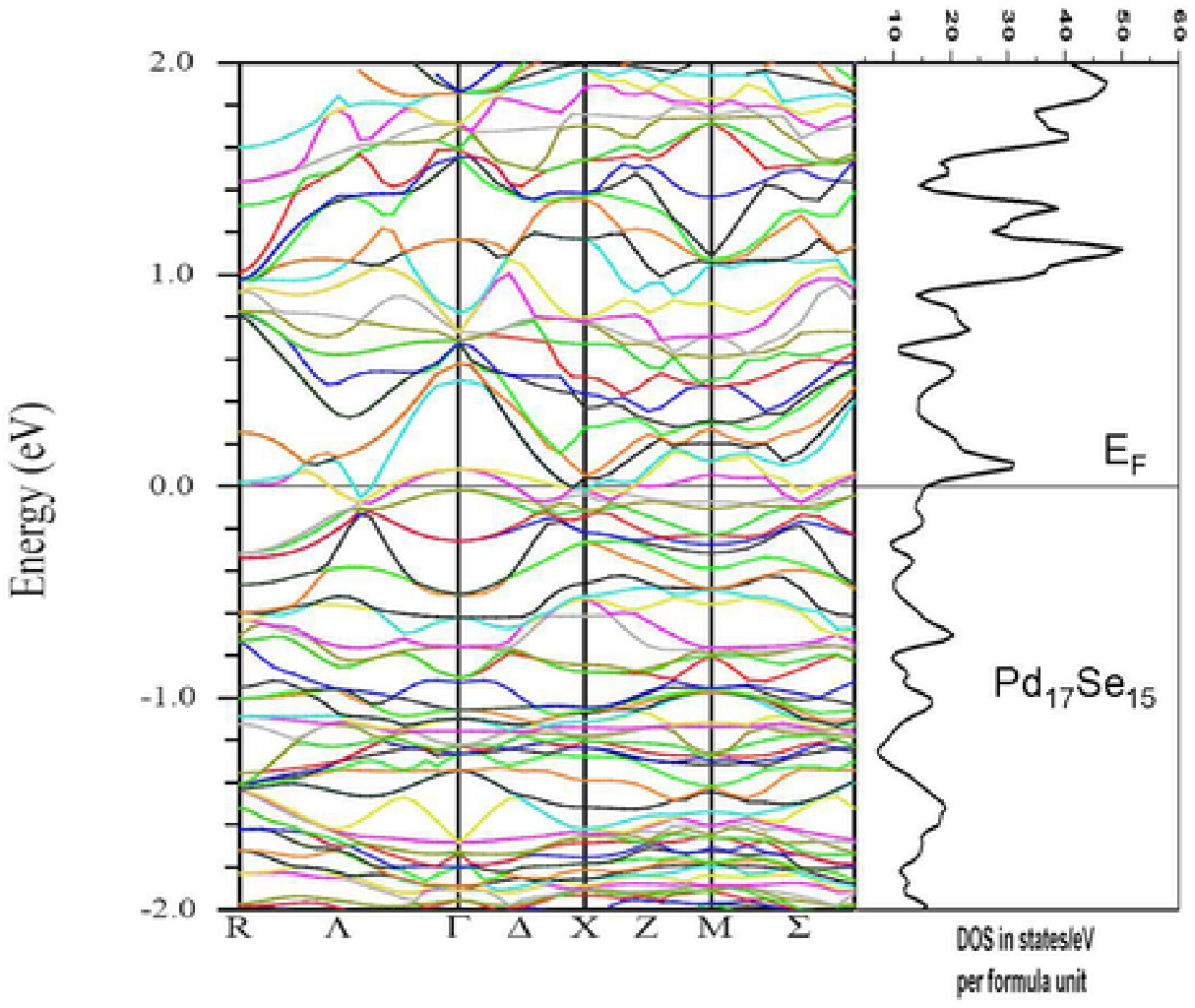}
	\caption{{[colour online] The band structure of Pd$_{17}$Se$_{15}$. The total DOS is indicated on the right side of the plot}}
	\label{PdSebs}
\end{figure*}

In fig. \ref{PdSebs} we have the band structure of Pd$_{17}$Se$_{15}$ juxtaposed with the DOS plot. We see that the Fermi level doesn't lie on a peak unlike Rh$_{17}$S$_{15}$. There is also, unlike Rh$_{17}$S$_{15}$, an absence of flat bands at Fermi level which which could be why the Fermi level  doesn't lie on a peak of DOS.

\begin{figure*}
	\centering
		\includegraphics[width=12cm]{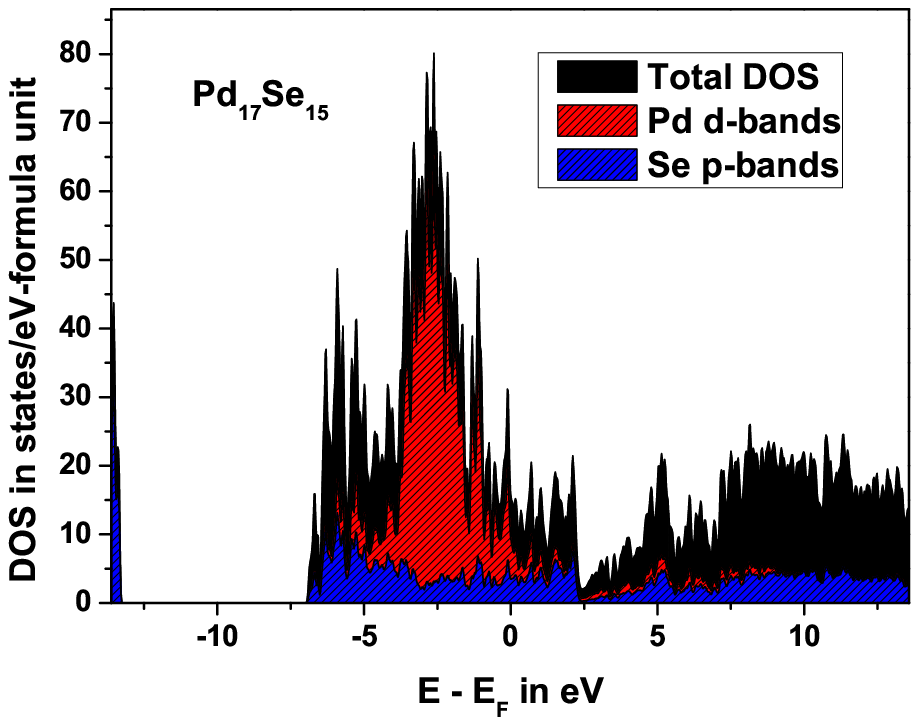}
	\caption{{[colour online] DOS for Pd$_{17}$Se$_{15}$ from LDA calculations. Contributions of Pd and Se are indicated in red and blue respectively}}
	\label{PdSeDOS}
\end{figure*}

Fig. \ref{PdSeDOS} has the DOS plot in the entire range of calcualation. Here, as in Rh$_{17}$S$_{15}$, the Pd d bands contribute more than 70 percent of the DOS at Fermi level. However, though the Pd 24m atoms are the major contributors to the total DOS, the Pd 1b atom's contribution per atom is larger than all the other Pd and Se atoms. We see that the DOS at Fermi level is around 16 states/eV-formula unit. This is a remarkably large value in comparison with that from the heat capacity data (which is around 9.3 states/eV-formula unit). This is a puzzling discrepancy since the enhanced DOS from heat capacity data is always expected to be larger than that obtained from an LDA calculation. To check if strong correlations are the origin of this discrepancy we performed LDA + U calculations on both the systems. 

\subsection{LDA + U calculations}
When strong correlations were taken into account in the calculations, the DOS at Fermi level decrease for both the compounds. The Fermi level still lies on a peak for Rh$_{17}$S$_{15}$ and in a valley for Pd$_{17}$Se$_{15}$. But the value of DOS at Fermi level is around 6 states/eV-formula unit for Pd$_{17}$Se$_{15}$ and 11.4 states/eV-formula unit for Rh$_{17}$S$_{15}$. This, when compared with the heat capacity estimates gives a $\lambda$ of 0.56 for Pd$_{17}$Se$_{15}$ and 4 for Rh$_{17}$S$_{15}$. While the $\lambda$ seems to match a McMillan's equation like estimate for Pd$_{17}$Se$_{15}$, it seems unrealistically large for Rh$_{17}$S$_{15}$. We comment on these calculations in the next section.

\section{Conclusions}
Earlier (Ref.~\cite{r1}), we had conjectured that the strong correlations, as deduced from the Sommerfeld coefficient, enhanced susceptibility and large upper critical field, in Rh$_{17}$S$_{15}$ were due to strong Rh-Rh interactions which form a large density of states at the Fermi level. On the other hand Pd$_{17}$Se$_{15}$, whose superconductivity we report for the first time here, has much weaker electron correlations as deduced from smaller values of the same parameters as mentioned above. Since the unit cell volume of Pd$_{17}$Se$_{15}$ is larger than Rh$_{17}$S$_{15}$, the interactions between Pd atoms could be weaker here and hence the weaker correlations.   

The deductions from the band structure calculations, however, are not straightforward. It appears that an LDA calculation seems to suffice to estimate the bare density of states of Rh$_{17}$S$_{15}$ while an LDA+U calculation seems necessary for Pd$_{17}$Se$_{15}$ to match the $\lambda$ values got from McMillian's equation. On the other hand LDA for Pd$_{17}$Se$_{15}$ and LDA+U for Rh$_{17}$S$_{15}$ seems to compare very badly with $\lambda$ from McMillian's equation. We believe that McMillian's equation, which is a based on a strong coupling BCS theory, is a reasonable estimate for Pd$_{17}$Se$_{15}$ since it is a conventional superconductor. Whereas McMillian's equation may not be applicable to Rh$_{17}$S$_{15}$ since it is an unconventional superconductor. This was also seen when the coupling parameter derived from the jump at the transition was much larger than $\lambda$ from McMillian's equation. As we saw from the upper critical field, it would require a $\lambda$ much larger than 3 to be Pauli limited in the BCS sense which is unrealistic. We believe that an LDA+U calculation is necessary to estimate the DOS of these systems and this gives a good estimate for Pd$_{17}$Se$_{15}$. However, DOS from an LDA+U calculation for Rh$_{17}$S$_{15}$ and its comparision with the DOS from heat capacity, cannot give a $\lambda$ matching a Mcmillan's equation estimate since Rh$_{17}$S$_{15}$ is a strongly correlated system.
It will interesting to study other families of chalcogenide superconductors with narrow 4d and 5d bands and compare for similarities with the '17-15' family, (Rh$_{17}$S$_{15}$ and Pd$_{17}$Se$_{15}$).

\section*{References}

\begin{thebibliography}{18}
\bibitem{r1} H.R. Naren, A. Thamizhavel, A. K. Nigam, and S. Ramakrishnan 2008 {\it Phys. Rev. Lett.} {\bf  100} 026404.
\bibitem{r2} H. R. Naren, A. Thamizhavel, S. Ramakrishnan and A. K. Grover 2009  {\it J. Phys.: Conf. Ser.} {\bf 150}  052183. 
\bibitem{r3} Sandip Dey and Vimal K Jain 2004 {\it Platinum Metals Review (Published by Johnson Matthey, Hatton Garden, London}) {\bf 48} 16.
\bibitem{r4} Oswaldo Diéguez and Nicola Marzari 2009 {\it Phys. Rev. B} {\bf 80} 214115.
\bibitem{r5} W.L. McMillan 1968 {\it Phys. Rev.} {\bf 167} 331.
\bibitem{r6} M. Uhlarz, O. Ignatchik, J. Wosnitza, H. R. Naren, A. Thamizhavel and S. Ramakrishnan 2010 {\it J. Low Temp. Phys.} {\bf 159} 176.
\bibitem{r7} Rikio Settai, Keisuke Katayama, Hiroshi Muranaka, Tetsuya Takeuchi, Arumugam Thamizhavel, Ilya Sheikin and Yoshisicko Onuki 2010 {\it Physics and Chemistry of Solids} {\bf 71} 700.
\bibitem{r8} N.R. Werthamer, E. Helfand and P. C. Hohenberg 1966 {\it Phys. Rev.} {\bf 147} 295.
\bibitem{r9} O. Fischer, 1978 {\it App. Phys. A} {\bf 16} 1.
\bibitem{r10} Article under preparation.
\bibitem{r11} G. Fuchs, K. -H. Müller, J. Freudenberger, K. Nenkov, S. L. Drechsler, S. V. Shulga, D. Lipp, A. Gladun, T. Cichorek and P. Gegenwart, 2002, {\it Pramana} {\bf 58} 791.
\bibitem{r12} K.-H. Müller, G. Fuchs, A. Handstein, K. Nenkov, V.N. Narozhnyi and D. Eckert, {\it arxiv.org/pdf/cond-mat/0102517}
\bibitem{r13}  R. W. Cohen, G. D. Cody, and J. J. Halloran, 1967 {\it Phys. Rev. Lett.} {\bf 19} 840.  
\bibitem{r14} J. Labbe 1967 {\it Phys. Rev.} {\bf 158} 647.
\bibitem{r15} T. Koyama, K. Kanda, K. Ueda, T. Mito, T. Kohara and H. Nakamura, 2010 {\it J. Phys.: Conf. Ser.} {\bf 200} 012095.
\bibitem{r16} P. Blaha, K. Schwarz, G. Madsen, D. Kvasnicka and J. Luitz, 2001 {\it www.wien2k.at}.
\bibitem{r17} K. Koepernik and H. Eschrig, 1999 {\it Phys. Rev. B} {\bf 59} 1743.
\bibitem{r18} I. Opahle, K. Koepernik and H. Eschrig, 1999 {\it Phys. Rev. B} {\bf 60} 14035.
\bibitem{r19} F. Hardy, T. Wolf, R. A. Fisher, R. Eder, P. Schweiss, P. Adelmann, H. v. Lohneysen and C. Meingast, 2010 {\it Phys. Rev. B} {\bf 81} 060501(R).
\end{thebibliography}
\end{document}